\documentclass[twocolumn, aps, pra, superscriptaddress, floatfix]{revtex4}
\usepackage[T1]{fontenc}
\usepackage[latin9]{inputenc}
\usepackage{color}
\usepackage{bm}% bold math
\usepackage{xcolor}
\usepackage{amsmath}
\usepackage{amssymb}
\usepackage{mathtools}
\usepackage{graphicx}
\usepackage[sort&compress]{natbib}
\usepackage{braket}
\usepackage{psfrag}
\usepackage{tikz}
\usepackage[normalem]{ulem}
\usepackage{qcircuit}
\usepackage{subcaption}
\usepackage{accents}
\usepackage{cleveref}
\usepackage{soul}
\usepackage{algorithm}
\usepackage{algpseudocode}
\usepackage{multirow}
\usepackage{makecell}

\newcommand{\varqbm}{VarQBM}

\begin{document}

\title{Variational Quantum Boltzmann Machines}% Force line breaks 

\author{Christa Zoufal}%
\affiliation{IBM Quantum, IBM Research -- Zurich}%, Rueschlikon 8803, Switzerland}
\affiliation{ETH Zurich}% 8092, Switzerland}

\author{Aur\'{e}lien Lucchi}
\affiliation{ETH Zurich}% 8092, Switzerland}

\author{Stefan Woerner}
\email{wor@zurich.ibm.com}
\affiliation{IBM Quantum, IBM Research -- Zurich}%, Rueschlikon 8803, Switzerland}
		
\date{\today}

\begin{abstract}
This work presents a novel realization approach to Quantum Boltzmann Machines (QBMs). 
The preparation of the required Gibbs states, as well as the evaluation of the loss function's analytic gradient is based on Variational Quantum Imaginary Time Evolution, a technique that is typically used for ground state computation. 
In contrast to existing methods, this implementation facilitates near-term compatible QBM training with gradients of the actual loss function for arbitrary parameterized Hamiltonians which do not necessarily have to be fully-visible but may also include hidden units.
The variational Gibbs state approximation is demonstrated with numerical simulations and experiments run on real quantum hardware provided by IBM Quantum.
Furthermore, we illustrate the application of this variational QBM approach to generative and discriminative learning tasks using numerical simulation.
\end{abstract}

\maketitle

%%%%%%%%%%%%%%%%%%%%%%%%%%%%%%%%%%%%%%%%%%%%%%%%%%%%%%%%%%%%%%%%%%%%%%%%%%
\section{Introduction}
%%%%%%%%%%%%%%%%%%%%%%%%%%%%%%%%%%%%%%%%%%%%%%%%%%%%%%%%%%%%%%%%%%%%%%%%%%

Boltzmann Machines (BMs) \cite{HintonBM1985, Du2019BM} offer a powerful framework for modelling probability distributions. 
These types of neural networks use an undirected graph-structure to encode relevant information. 
More precisely, the respective information is stored in bias coefficients and connection weights of network nodes, which are typically related to binary spin-systems and grouped into those that determine the output, the visible nodes, and those that act as latent variables, the hidden nodes.
Furthermore, the network structure is linked to an energy function which facilitates the definition of a probability distribution over the possible node configurations by using a concept from statistical mechanics, i.e., Gibbs states \cite{Boltzmann1877, gibbs02}.
The aim of BM training is to learn a set of weights such that the resulting model approximates a target probability distribution which is implicitly given by training data. 
This setting can be formulated as discriminative as well as generative learning task \cite{Liu2010}.
Applications have been studied in a large variety of domains such as the analysis of quantum many-body systems, statistics, biochemistry, social networks, signal processing and finance, see, e.g., \cite{CarleoRBMsQManyBody18, CarleoRBMsQuantumManyBody17, YusukeRBM17, AnshuSample-efficientQManyBody, Melko2019, HRASKO2015RBMTimeSeries, Tubiana19RBMProteins, LiuRBMsSocialNetworks13, Mohamed10RBMSignal, Assis18RBMFin}.
However, BMs are complicated to train in practice because the loss function's derivative requires the evaluation of a normalization factor, the partition function, that is generally difficult to compute.
Usually, it is approximated using Markov Chain Monte Carlo methods which may require long runtimes until convergence \cite{Hinton05CD, MurphyML12}. Alternatively, the gradients could be estimated approximately using contrastive divergence \cite{Hinton2002TrainingPO} or pseudo-likelihood \cite{Besag1975} potentially leading to inaccurate results \cite{Tieleman08, Sutskever10}.

%Existing work
Quantum Boltzmann Machines (QBMs) \cite{QBMAmin18} are a natural adaption of BMs to the quantum computing framework. Instead of an energy function with nodes being represented by binary spin values, QBMs define the underlying network using a Hermitian operator, a parameterized Hamiltonian
\begin{equation*}
    H_{\theta}=\sum_{i=0}^{p-1}\theta_ih_i,
\end{equation*}
with $\theta\in\mathbb{R}^p$ and $h_i=\bigotimes_{j=0}^{n-1}\sigma_{j, i}$ for $\sigma_{j, i}\in\set{I, X, Y, Z}$ acting on the $j^{\text{th}}$ qubit. 
The network nodes are hereby characterized by the Pauli matrices $\sigma_{j, i}$.
This Hamiltonian relates to a quantum Gibbs state, $\rho^{\text{Gibbs}} = {e^{-H_{\theta}/\left(\text{k}_{\text{B}}\text{T}\right)}}/{Z}$
with $\text{k}_{\text{B}}$ and $\text{T}$ denoting the Boltzmann constant and the system temperature, and  $Z=\text{Tr}\left[e^{-H_{\theta}/\left(\text{k}_{\text{B}}\text{T}\right)}\right]$. 
It should be noted that those qubits which determine the model output are referred to as visible and those which act as latent variables as hidden qubits.
The aim of the model is to learn Hamiltonian parameters such that the resulting Gibbs state reflects a given target system.
In contrast to BMs, this framework allows the use of quantum structures which are potentially inaccessible classically.
Equivalently to the classical model, QBMs are suitable for discriminative as well as generative learning.

We present here a QBM implementation that circumvents certain issues which emerged in former approaches.
The first paper on QBMs \cite{QBMAmin18} and several subsequent works \cite{Anschtz2019RealizingQB, QBMWiebe17, Kappen18QBM, Wiebe2019GenerativeTO} are incompatible with efficient evaluation of the loss function's analytic gradients if the given model has hidden qubits and 
\begin{equation*}
    \exists j: \:\left[H_{\theta}, \frac{\partial H_{\theta}}{\partial\theta_j}\right] \neq 0.
\end{equation*}
Instead, the use of hidden qubits is either avoided, i.e., only fully-visible settings are considered \cite{QBMWiebe17, Kappen18QBM, Wiebe2019GenerativeTO}, or the gradients are computed with respect to an upper bound of the loss \cite{QBMAmin18, Anschtz2019RealizingQB, QBMWiebe17}, which is based on the Golden-Thompson inequality \cite{ThompsonInequality1965, Golden65Helm}.
It should be noted that training with an upper bound, renders the use of transverse Hamiltonian components, i.e., off-diagonal Pauli terms, difficult and imposes restrictions on the compatible models.

Further, we would like to point out that, in general, it is not trivial to evaluate a QBM Hamiltonian with a classical computer, i.e., using exact simulation with Quantum Monte Carlo methods \cite{Troyer05}, because the underlying Hamiltonian can suffer from the so-called \emph{sign-problem} \cite{Hangleiter2019EasingTM, OkunishiSignProblem14, Li2016SignProblemFreeQMC, SignProblemAlet16, PhysRevB.91.241117}. As already discussed in \cite{OrtizQAFermionic01}, evaluations on quantum computers can avoid this problem.

%Our QBM approach
Our QBM implementation works for generic Hamiltonians $H_{\theta}$ with real coefficients $\theta$ and arbitrary Pauli terms $h_i$, and furthermore, is compatible with near-term, gate-based quantum computers.
The method exploits \emph{Variational Quantum Imaginary Time Evolution} \cite{VarSITEMcArdle19, Simon18TheoryVarQSim} (VarQITE), which is based on McLachlan's variational principle \cite{McLachlan64}, to not only prepare approximate Gibbs states, $\rho_{\omega}^{\text{Gibbs}}$, but also to train the model with gradients of the actual loss function.
During each step of the training, we use VarQITE to generate an approximation to the Gibbs state underlying $H_{\theta}$ and to enable automatic differentiation for computing the gradient of the loss function which is needed to update $\theta$.
This \emph{Variational QBM} algorithm (\varqbm) is inherently normalized which implies that the training does not require the explicit evaluation of the partition function. 

We focus on training quantum Gibbs states whose sampling behavior reflects a classical probability distribution. However, the scheme could be easily adapted to an approximate quantum state preparation scheme by using a loss function which is based on the quantum relative entropy \cite{QBMWiebe17, Kappen18QBM, Wiebe2019GenerativeTO}.
Hereby, the approximation to $\rho^{\text{Gibbs}}$ is fitted to a given target state $\rho^{\text{data}}$. 
Notably, this approach is not necessarily suitable for learning classical distributions. More precisely, we do not need to train a quantum state that captures all features of the density matrix $\rho^{\text{data}}$ but only those which determine the sampling probability. 
It follows that fitting the full density matrix may impede the training.

The remainder of this paper is structured as follows. Firstly, we review classical BMs and VarQITE in Sec.~\ref{sec:pre}.
Then, we outline \varqbm{} in Sec.~\ref{sec:QBM}. Next, we illustrate the feasibility of the Gibbs state preparation and present QBM applications in Sec.~\ref{sec:results}.
Finally, a conclusion and an outlook are given in Sec.~\ref{sec:discussion}.

%%%%%%%%%%%%%%%%%%%%%%%%%%%%%%%%%%%%%%%%%%%%%%%%%%%%%%%%%%%%%%%%%%%%%%%%%%
\section{Preliminaries}
\label{sec:pre}
%%%%%%%%%%%%%%%%%%%%%%%%%%%%%%%%%%%%%%%%%%%%%%%%%%%%%%%%%%%%%%%%%%%%%%%%%%

This section introduces the concepts which form the basis of our \varqbm{} algorithm. First, classical BMs are presented in Sec.~\ref{sec:classBM}. Then, we discuss VarQITE, the algorithm that \varqbm{} uses for approximate Gibbs state preparation, in Sec.~\ref{sec:VarQITE}.

%%%%%%%%%%%%%%%%%%%%%%%%%%%%%%%%%%%%%%%%%%%%%%%%%%%%%%
\subsection{Boltzmann Machines}
\label{sec:classBM}
%%%%%%%%%%%%%%%%%%%%%%%%%%%%%%%%%%%%%%%%%%%%%%%%%%%%%%%%%%%%%%%%%%%%%

Here, we will briefly review the original concept of classical BMs \cite{HintonBM1985}.
A BM represents a network model that stores the learned knowledge in connection weights between network nodes. 
More explicitly, the connection weights are trained to generate outcomes according to a probability distribution of interest, e.g., to generate samples which are similar to given training samples or to output correct labels depending on input data samples. 

Typically, this type of neural network is related to an Ising-type model \cite{Ising1925, peierls_1936} such that each node $i$ corresponds to a binary variable $z_i \in \set{-1, +1}$.
Now, the set of nodes may be split into visible and hidden nodes representing observed and latent variables, respectively.
Furthermore, a certain configuration $z=\left\{v,\: h\right\}$ of all nodes -- visible and hidden -- determines an energy, which is given as 
\begin{equation*}
    E_{z = \left\{v,\: h\right\}}= -\sum\limits_i\tilde{\theta}_iz_i - \sum\limits_{i, j}\theta_{ij}z_iz_j,
\end{equation*}
with $\tilde{\theta}_i, \theta_{ij}\in\mathbb{R}$ denoting the weights and $z_i$ representing the value taken by node $i$. It should be noted that the parameters $\theta_{ij}$ correspond to the weights of connections between different nodes. More explicitly, if two nodes are connected in the network, then a respective term appears in the energy function.
The probability to observe a configuration $v$ of the visible nodes is defined as \begin{equation}
\label{eq:gibbs_dis}
    p^{BM}_v = \frac{e^{-E_v/\left(\text{k}_{\text{B}}\text{T}\right)}}{Z},
\end{equation}
where $E_v = \sum_h E_{z = \left\{v,\: h\right\}}$, $k_B$ is the Boltzmann constant, $T$ the system temperature and $Z$ the canonical partition function
\begin{equation*}
    Z=\sum\limits_{z= \left\{v,\: h\right\}}e^{-E_z/\left(\text{k}_{\text{B}}\text{T}\right)}.
\end{equation*}
We would like to point out that BMs adopt a concept from statistical mechanics.
Suppose a closed system that is in thermal equilibrium with a coupled heat bath at constant temperature. The possible configuration space is determined by the canonical ensemble, i.e., the probability for observing a configuration is given by the Gibbs distribution \cite{Boltzmann1877, gibbs02} which corresponds to Eq.~\eqref{eq:gibbs_dis}.

Now, the goal of a BM is to fit the target probability distribution $p^{\text{data}}$ with $p^{BM}$.
Typically, this training objective is achieved by optimizing the cross-entropy
\begin{equation}
\label{eq:crossEnt}
	L = -\sum\limits_{v}p_v^{\text{data}}\log{p_v^{BM}}.
\end{equation}
In theory, fully-connected BMs have interesting representation capabilities \cite{HintonBM1985, YOUNES1996109, Fischer12RBM}, i.e., they are universal approximators \cite{Roux10}. 
However, in practice they are difficult to train as the optimization easily gets expensive. 
Thus, it has become common practice to restrict the connectivity between nodes which relates to restricted Boltzmann Machines (RBMs) \cite{RBM_Montufar_2018}. 
Furthermore, several approximation techniques, such as contrastive divergence \cite{Hinton2002TrainingPO}, have been developed to facilitate BM training. 
However, these approximation techniques typically still face issues such as long computation time due to a large amount of required Markov chain steps or poor compatibility with multimodal probability distributions \cite{MurphyML12}.
For further details, we refer the interested reader to \cite{Hinton2012, Fischer2012, Fischer2015}.

%%%%%%%%%%%%%%%%%%%%%%%%%%%%%%%%%%%%%%%%%%%%%%%%%%%%%%%%%%%%%%%%%%%%%
\subsection{Variational Quantum Imaginary Time Evolution}
\label{sec:VarQITE}
%%%%%%%%%%%%%%%%%%%%%%%%%%%%%%%%%%%%%%%%%%%%%%%%%%%%%%%%%%%%%%%%%%%%%

Imaginary time evolution (ITE) \cite{MagnusITE54} is an approach that is well known for (classical) ground state computation \cite{VarSITEMcArdle19, GuptaITE02, ITEAuer01}. 

Suppose a starting state $\ket{\psi_0}$ and a time-independent Hamiltonian $H=\sum_{i=0}^{p-1}\theta_ih_i$ with real coefficients $\theta_i$ and Pauli terms $h_i$. Then, the normalized ITE propagates $\ket{\psi_0}$ with respect to $H$ for time $\tau$ according to
\begin{equation*}
\label{eq:ITE}
	\ket{\psi_\tau}=  C\left(\tau\right)e^{-H\tau}\ket{\psi_0},
\end{equation*}
where $C\left(\tau\right) = {1}/{\sqrt{\text{Tr}\left[e^{-2H\tau}\ket{\psi_0 }\bra{\psi_0}\right]}}$ is a normalization.
The differential equation that describes this evolution is the Wick-rotated Schr\"odinger equation
\begin{equation}
\label{eq:WickSchroed}
	\frac{d\ket{\psi_\tau}}{d \tau} =-\left( H - E_{\tau} \right) \ket{\psi_\tau}, 
\end{equation}
where $E_{\tau} = \bra{\psi_{\tau}} H\ket{\psi_\tau}$ originates from the normalization of $\ket{\psi_{\tau}}$.
The terms in $e^{-H\tau}$, corresponding to small eigenvalues of $H$, decay slower than the ones corresponding to large eigenvalues. Due to the continuous normalization, the smallest eigenvalue dominates for $\tau \rightarrow \infty$. Thus, $\ket{\psi_\tau}$ converges to the ground state of $H$ given that there is some overlap between the ground and starting state.
Furthermore, if ITE is only evolved to a finite time, $\tau = {1}/{2\left(\text{k}_{\text{B}}\text{T}\right)}$, then it enables the preparation of Gibbs states, see Sec.~\ref{sec:gibbsVarQITE}. 

As introduced in \cite{VarSITEMcArdle19, Simon18TheoryVarQSim}, an approximate ITE can be implemented on a gate-based quantum computer by using  McLachlan's variational principle \cite{McLachlan64}.
The basic idea of the method is to introduce a parameterized trial state $\ket{\psi_{\omega}}$ and to project the temporal evolution of $\ket{\psi_\tau}$ to the parameters, i.e.,  $\omega \coloneqq \omega(\tau)$.
We refer to this algorithm as VarQITE and, now, discuss it in more detail.

First, we define an input state $\ket{\psi_{\text{in}}}$ and a quantum circuit $V\left(\omega\right) = U_q\left(\omega_q\right)\cdots U_1\left(\omega_1\right)$ with parameters $\omega \in \mathbb{R}^{q}$ to generate the parameterized trial state
\begin{equation*}
\label{eq:varSite}
\ket{\psi_{\omega}} \coloneqq V\left(\omega\right)\ket{\psi_{\text{in}}}.
\end{equation*}
Now, McLachlan's variational principle 
\begin{equation}
\label{eq:McLachlan}
	\delta  \left\lVert \left(d/d\tau + H - E_{\tau}\right) \ket{ \psi_{\omega}} \right\rVert = 0
\end{equation}
determines the time propagation of the parameters $\omega(\tau)$.
This principle aims to minimize the distance between the right hand side of Eq.~\eqref{eq:WickSchroed} and the change $d\ket{ \psi_{\omega}} / d \tau$.
Eq.~\eqref{eq:McLachlan} leads to a system of linear equations for  $\dot{\omega}= d \omega / d\tau$, i.e., 
\begin{align}
\label{eq:sle}
A\dot{\omega} = C
\end{align}
with
\begin{equation}
\begin{split}
\label{eq:AC}
A_{pq}\left(\tau\right) &= \text{Re}\left(\text{Tr}\left[\frac{\partial V^{\dagger}\left(\omega\left(\tau\right)\right)}{\partial\omega\left(\tau\right)_p}\frac{\partial V\left(\omega\left(\tau\right)\right)}{\partial\omega\left(\tau\right)_q}\rho_{\text{in}}\right]\right) \\
C_p\left(\tau\right) &=  -\sum\limits_i\theta_i\text{Re}\left(\text{Tr}\left[\frac{\partial V^{\dagger}\left(\omega\left(\tau\right)\right)}{\partial\omega\left(\tau\right)_p}h_iV\left(\omega\left(\tau\right)\right)\rho_{\text{in}}\right]\right),
\end{split}
\end{equation}
where $\text{Re}\left(\cdot\right)$ denotes the real part and $\rho_{\text{in}} = \ket{\psi_{\text{in}}}\bra{\psi_{\text{in}}}$.
The vector $C$ describes the derivative of the system energy $ \bra{ \psi_{\omega}}H \ket{ \psi_{\omega}}$ and $A$ is proportional to the classical Fisher information matrix, a metric tensor that reflects the system's information geometry \cite{QNGNonUnitary19Simon}.
To evaluate $A$ and $C$, we compute expectation values with respect to quantum circuits of a particular form which is illustrated and discussed in Appendix \ref{app:a_c}. 

This evaluation is compatible with arbitrary parameterized unitaries in $V\left(\omega\right)$ because all unitaries can be written as $U\left(\omega\right) = e^{iM\left(\omega\right)}$, where $M\left(\omega\right)$ denotes a parameterized Hermitian matrix. 
Further, Hermitian matrices can be decomposed into weighted sums of Pauli terms, i.e., $M\left(\omega\right) = \sum_pm_p\left(\omega\right)h_p$ with $m_p\left(\omega\right)\in\mathbb{R}$ and $h_p=\bigotimes\limits_{j=0}^{n-1}\sigma_{j, p}$ for $\sigma_{j, p}\in\set{I, X, Y, Z}$ \cite{nielsen10} acting on the $j^{\text{th}}$ qubit. Thus, the gradients of 
$U_k\left(\omega_k\right)$ are given by
\begin{equation*}
\frac{\partial U_k\left(\omega_k\right)}{\partial\omega_k} = \sum\limits_pi \frac{\partial m_{k,p}\left(\omega_k\right)}{\partial\omega_k}U_k\left(\omega_k\right)h_{k_p}.
\end{equation*}
This decomposition allows us to compute $A$ and $C$ with the techniques described in \cite{LaflammeSimulatingPhysPhenom02, VarSITEMcArdle19, Simon18TheoryVarQSim}.
Furthermore, it should be noted that Eq.~\eqref{eq:sle} is often ill-conditioned and may, thus, require the use of regularized regression methods, see Sec.~\ref{sec:impRel}.
 
Now, we can use, e.g., an explicit Euler method to evolve the parameters as 
\begin{equation*}
\omega\left(\tau\right) = \omega\left(0\right) + \sum\limits_{j = 1}^{\tau/\delta\tau} \dot{\omega}\left(\tau\right)\delta\tau.
\end{equation*}

%%%%%%%%%%%%%%%%%%%%%%%%%%%%%%%%%%%%%%%%%%%%%%%%%%%%%%%%%%%%%%%%%%%%%
\section{Quantum Boltzmann Machine Algorithm}
\label{sec:QBM}
%%%%%%%%%%%%%%%%%%%%%%%%%%%%%%%%%%%%%%%%%%%%%%%%%%%%%%%%%%%%%%%%%%%%%%%%%%

A QBM is defined by a parameterized Hamiltonian $H_{\theta}=\sum_{i=0}^{p-1}\theta_ih_i $ where $\theta\in\mathbb{R}^p$ and $h_i=\bigotimes_{j=0}^{n-1}\sigma_{j, i}$ for $\sigma_{j, i}\in\set{I, X, Y, Z}$ acting on the $j^{\text{th}}$ qubit. 
Equivalently to classical BMs, QBMs are typically represented by an Ising model \cite{Ising1925}, i.e., a $2$-local system \cite{bravyi06LocalHam} with nearest-neighbor coupling that is defined with regard to a particular grid. In principle, however, any Hamiltonian compatible with Boltzmann distributions could be used. 

In contrast to BMs, the network nodes, given by the Pauli terms $\sigma_{j, i}$, do not represent the visible and hidden units. These are defined with respect to certain sub-sets of qubits. More explicitly, those qubits which determine the output of the QBM are the visible qubits, whereas the others correspond to the hidden qubits.
Now, the probability to measure a configuration $v$ of the visible qubits is defined with respect to a projective measurement $\Lambda_v = \ket{v}\bra{v}\otimes I$ on the quantum Gibbs state 
\begin{equation*}
\label{eq:QGibbs}
\rho^{\text{Gibbs}} = \frac{e^{-H_{\theta}/\left(\text{k}_{\text{B}}\text{T}\right)}}{Z}
\end{equation*}
with $Z=\text{Tr}\left[e^{-H_{\theta}/\left(\text{k}_{\text{B}}\text{T}\right)}\right]$, i.e., the probability to measure $\ket{v}$ is given by 
\begin{equation*}
    p_v^{\text{QBM}} = \text{Tr}\left[\Lambda_v\rho^{\text{Gibbs}} \right].
\end{equation*}
For the remainder of this work, we assume that $\Lambda_v$ refers to projective measurements with respect to the computational basis of the visible qubits. Thus, the configuration $v$ is determined by $v_i\in \set{0, 1}$. 
It should be noted that this formulation does not require the evaluation of the configuration of the hidden qubits.

Our goal is to train the Hamiltonian parameters $\theta$ such that the sampling probabilities of the corresponding $\rho^{\text{Gibbs}}$ reflect the probability distribution underlying given classical training data. For this purpose, the same loss function as described in the classical case, see Eq.~\eqref{eq:crossEnt}, can be used
\begin{equation}
\label{eq:loss_qbm}
	L = -\sum\limits_{v}p_v^{\text{data}}\log{p_v^{\text{QBM}}},
\end{equation}
where $p_v^{\text{data}}$ denotes the occurrence probability of item $v$ in the training data set.

\begin{figure}[h!]
\captionsetup{singlelinecheck = false, format= hang, justification=raggedright, font=footnotesize, labelsep=space}
\begin{center}
\includegraphics[width=0.8\linewidth]{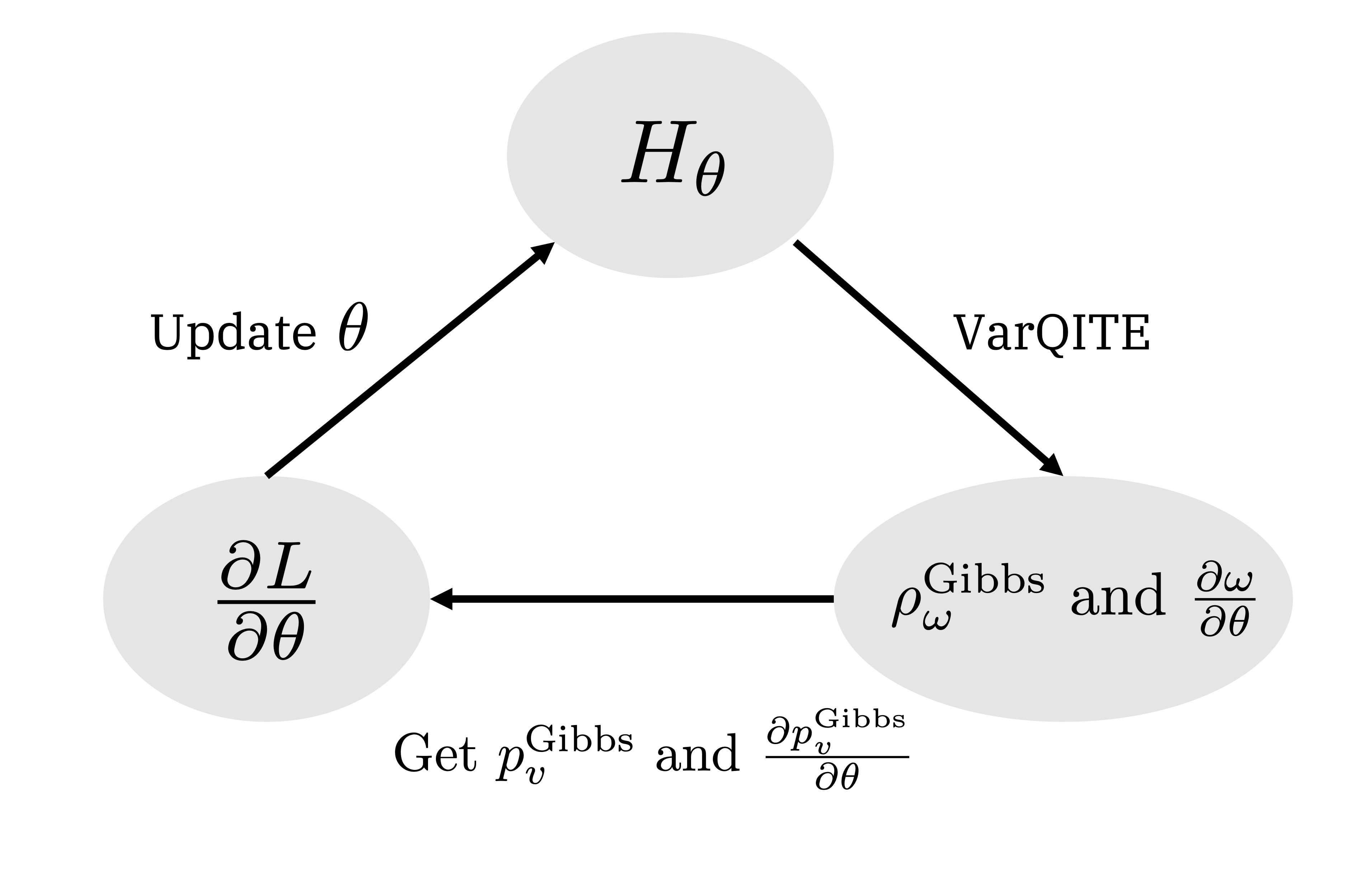}
\end{center}
\caption{The \varqbm{} training includes the following steps. First, we need to fix the Pauli terms for $H_{\theta}$ and choose initial parameters $\theta$. Then, VarQITE is used to generate $\rho_{\omega}^{\text{Gibbs}}$ and compute $\partial\omega/ \partial\theta$. 
The quantum state and the derivative are needed to evaluate $p_v^{\text{QBM}}$ and $\partial p_v^{\text{QBM}}/\partial\theta$. Now, we can find $\partial L/\partial\theta$ to update the Hamiltonian parameters with a classical optimizer.}
\label{fig:varqbm}
\end{figure}

To enable efficient training, we want to evaluate the derivative of $L$ with respect to the Hamiltonian parameters. Unlike existing QBM implementations, \varqbm{} facilitates the use of analytic gradients of the loss function given in Eq.~\eqref{eq:loss_qbm} for generic QBMs.
The presented algorithm involves the following steps.
First, we use VarQITE to approximate the Gibbs state, see Sec.~\ref{sec:gibbsVarQITE} for further details. 
Then, we compute the gradient of $L$  to update the parameters $\theta$ with automatic differentiation, as is discussed in Sec.~\ref{sec:implementation}. The parameters are trained with a classical optimization routine where one training step consists of the Gibbs state preparation with respect to the current parameter values and a consecutive parameter update, as illustrated in Fig.~\ref{fig:varqbm}.

In the remainder of this section, we discuss Gibbs state preparation with VarQITE in Sec.~\ref{sec:gibbsVarQITE} and VarQBM in more detail in Sec.~\ref{sec:implementation}.

%%%%%%%%%%%%%%%%%%%%%%%%%%%%%%%%%%%%%%%%%%%%%%%%%%%%%%%%%%%%%%%%%%%%%
\subsection{Gibbs State Preparation with VarQITE}
\label{sec:gibbsVarQITE}
%%%%%%%%%%%%%%%%%%%%%%%%%%%%%%%%%%%%%%%%%%%%%%%%%%%%%%%%%%%%%%%%%%%%%

The Gibbs state $\rho^{\text{Gibbs}}$ describes the probability density operator of the configuration space of a system in thermal equilibrium with a heat bath at constant temperature $T$ \cite{gibbs_2010}. Originally, Gibbs states were studied in the context of statistical mechanics but, as shown in \cite{Pauli1927}, the density operator also facilitates the description of quantum statistics.
 
Gibbs state preparation can be approached from different angles. Hereby, different techniques not only have different strengths but also different drawbacks. 
Some schemes \cite{Temme2011QuantumMS, YungQuantumMetropolis12, PoulinThermalQGibbs09} use Quantum Phase Estimation \cite{AbramsQPE99} as a subroutine, which is likely to require error-corrected quantum computers. 
Other methods enable the evaluation of quantum thermal averages \cite{MottaQITE20, brandaoFiniteCorrLengthEfficientPrep19, BrandaoGibbsSampler16} for states with finite correlations. However, since QBM-related states may exhibit long-range correlations, these methods are not the first choice for the respective preparation.
A thermalization based approach is presented in \cite{Anschtz2019RealizingQB}, where the aim is to prepare a quantum Gibbs state by coupling the state register to a heat bath given in the form of an ancillary quantum register. 
Correct preparation requires a thorough study of suitable ancillary registers for a generic Hamiltonian as the most useful ancilla system is not a-priori known.
Further, a variational Gibbs state preparation method has been presented \cite{WiebeVariationalGibbs2020} which is based on the fact that Gibbs states minimize the free energy of a system at constant temperature.
Thus, the goal is to fit a parameterized quantum state such that it minimizes the free energy. 
The parameter update is hereby conducted with a finite difference method instead of analytic gradients which may impair the training accuracy. Additionally, the method requires the application of Quantum Amplitude Estimation \cite{brassardQAE02}, as well as matrix exponentiation of the input state, and thus, is not well suited for near-term quantum computing applications.

In contrast to these Gibbs state preparation schemes, VarQITE is compatible with near-term quantum computers, and is neither limited to states with finite correlations nor requires ambiguous ancillary systems.
In the following, we discuss how VarQITE can be utilized to generate an approximation of the Gibbs state $\rho^{\text{Gibbs}}$ for a generic $n-$qubit Hamiltonian $H_{\theta}=\sum_{i=0}^{p-1}\theta_ih_i $ with $\theta\in\mathbb{R}^p$ and $h_i=\bigotimes_{j=0}^{n-1}\sigma_{j, i}$ for $\sigma_{j, i}\in\set{I, X, Y, Z}$ acting on the $j_i^{\text{th}}$ qubit.

First, we need to choose a suitable variational quantum circuit $V\left(\omega\right)$, $\omega \in \mathbb{R}^{q}$, and set of initial parameters $\omega(0)$ such that the initial state is
\begin{equation*}
     \ket{ \psi_{0}} = V\left(\omega\left(0\right)\right)\ket{0}^{\otimes 2n} = \ket{\phi^{+}}^{\otimes n}
\end{equation*}
where $\ket{\phi^{+}} = \frac{1}{\sqrt{2}}\left(\ket{00}+\ket{11}\right)$ represents a Bell state.
We define two $n$-qubit sub-systems $a$ and $b$ such that the first and second qubit of each $\ket{\phi^{+}}$ is in $a$ and $b$, respectively. Accordingly, an effective $2n$-qubit Hamiltonian $H_{\text{eff}} = H_{\theta}^a + I^b$, where $H_{\theta}$ and $I$ act on sub-system $a$ and $b$, is considered.
It should be noted that tracing out sub-system $b$ from $\ket{ \psi_{0}}$ results in an $n$-dimensional maximally mixed state
\begin{equation*}
		  \text{Tr}_{b}\left[\ket{\phi^{+}}^{\otimes n} \right] = \frac{1}{2^n}I.
\end{equation*}
Now, the Gibbs state approximation $\rho_{\omega}^{\text{Gibbs}}$ can be generated by propagating the trial state with VarQITE with respect to $H_{\text{eff}}$ for $\tau = 1/2\left(\text{k}_{\text{B}}\text{T}\right)$.
The resulting state
\begin{equation*}
	\ket{ \psi_{\omega}} = V\left(\omega\left(\tau\right)\right)\ket{0}^{\otimes 2n}
\end{equation*}
gives an approximation for the Gibbs state of interest
\begin{equation*}	
	\rho_{\omega}^{\text{Gibbs}} = \text{Tr}_{b}\left[\ket{ \psi\left({\omega\left(\tau\right)}\right)}\bra{ \psi\left({\omega\left(\tau\right)}\right)} \right] \approx  \frac{e^{-H_{\theta}/\left(\text{k}_{\text{B}}\text{T}\right)}}{Z}
\end{equation*}
by tracing out the ancillary system $b$.
We would like to point out that the VarQITE propagation relates $\omega$ to $\theta$ via the energy derivative $C$ given in Eq.~\eqref{eq:AC}.

Equivalently to Eq.~\eqref{eq:sle}, Eq.~\eqref{eq:sle_derivative} is also prone to being ill-conditioned. Thus, the use of regularization schemes may be required.

Notably, this is an approximate state preparation scheme that relies on the representation capabilities of $\ket{ \psi_{\omega}}$. However, since the algorithm is employed in the context of machine learning we do not necessarily require perfect state preparation. The noise may even improve the training, as discussed e.g., in \cite{Noh2017RegularizingDN}.

McLachlan's variational principle is not only the key component for Gibbs state preparation. It also enables the QBM training with gradients of the actual loss function for generic Pauli terms in $H_{\theta}$, even if some of the qubits are hidden. Further details are given in Sec.~\ref{sec:implementation}.

%%%%%%%%%%%%%%%%%%%%%%%%%%%%%%%%%%%%%%%%%%%%%%%%%%%%%%%%%%%%%%%%%%%%%
\subsection{Variational QBM}
\label{sec:implementation}
%%%%%%%%%%%%%%%%%%%%%%%%%%%%%%%%%%%%%%%%%%%%%%%%%%%%%%%%%%%%%%%%%%%%%

In the following, \varqbm{} and the respective utilization of McLachlan's variational principle and VarQITE is discussed.
We consider training data that takes at most $2^n$ different values and is distributed according to a discrete probability distribution $p^{\text{data}}$.
The aim of a QBM is to train the parameters of $H_{\theta}$ such that the sampling probability distribution of the corresponding $\rho_{\omega}^{\text{Gibbs}}= e^{-H_{\theta}/\left(\text{k}_{\text{B}}\text{T}\right)} / Z$ for $\ket{v}, \: v\in{0, \ldots, 2^n-1}$ with
\begin{equation*}
 p_v^{\text{QBM}} = \text{Tr}\left[\Lambda_v\rho_{\omega}^{\text{Gibbs}}\right],    
\end{equation*}
approximates $p^{\text{data}}$.
The QBM model is trained to represent $p^{\text{data}}$ by minimizing the loss, given in Eq.~\eqref{eq:loss_qbm}, with respect to the Hamiltonian parameters $\theta$, i.e.,
\begin{equation*}
    \underset{\theta}{\min} \: L = \underset{\theta}{\min} \: \left(-\sum\limits_{v}p_v^{\text{data}}\log{p_v^{\text{QBM}}}\right).
\end{equation*}
Now, \varqbm{} facilitates gradient-based optimization with the derivative of the actual loss function
\begin{equation}
\label{eq:loss_derivative}
\begin{split}
	\frac{\partial L}{\partial\theta_i} &= \frac{\partial\left( - \sum\limits_{v}p_v^{\text{data}}\log{p_v^{\text{QBM}}}\right)}{\partial\theta_i} \\ &= - \sum\limits_{v}p_v^{\text{data}}\frac{\partial p_v^{\text{QBM}}/\partial\theta_i}{p_v^{\text{QBM}}}
		\end{split}
\end{equation}
by using the chain rule, i.e., automatic differentiation.
More precisely, the gradient of $L$  can be computed by using the chain rule for
\begin{align}
\label{eq:gradGibbsState}
\begin{split}
	\frac{\partial p_v^{\text{QBM}}}{\partial\theta_i}  &=  
	 \frac{\partial  p_v^{\text{QBM}}}{\partial\omega\left(\tau\right)}\frac{\partial\omega\left(\tau\right)}{\partial\theta_i} \\
	 &=\sum\limits_{k=0}^{q-1} \frac{\partial  p_v^{\text{QBM}} }{\partial\omega_k\left(\tau\right)}\frac{\partial\omega_k\left(\tau\right)}{\partial\theta_i}.
	 \end{split}
\end{align}
Firstly, $\partial  p_v^{\text{QBM}}/\partial\omega_k\left(\tau\right)=\partial \text{Tr}\left[\Lambda_v\rho_{\omega}^{\text{Gibbs}}\right] / \partial\omega_k\left(\tau\right)$ can be evaluated with quantum gradient methods discussed in \cite{Farhi2018_gradients, Fujii2018_qcircuitLearn,killoran2018, SchuldQuantumGradients19, Zoufal2019} because the term has the following form $\partial \text{Tr}\left[\hat{O}\ket{\phi\left(\alpha\right)}\bra{\phi\left(\alpha\right)}\right] / \partial\alpha$.
Secondly, ${\partial\omega_k\left(\tau\right)}/{\partial\theta_i}$ is evaluated by computing the derivative of Eq.~\eqref{eq:sle} with respect to the Hamiltonian parameters
\begin{equation*} 
				 \frac{\partial A\dot{\omega}\left(\tau\right)}{\partial{\theta_i}} = \frac{\partial  C}{\partial{\theta_i}}.
\end{equation*}
This gives the following system of linear equations
\begin{align}
\label {eq:sle_derivative}
         A\left(\frac{\partial\dot{\omega}\left(\tau\right)}{\partial{\theta_i}}\right) = \frac{\partial C}{\partial{\theta_i}} - \left(\frac{\partial A}{\partial{\theta_i}}\right)\dot{\omega}\left(\tau\right).
\end{align}
Now, solving for $\partial\dot{\omega}\left(\tau\right)/\partial{\theta_i}$ in every time step of the Gibbs state preparation enables the use of, e.g., an explicit Euler method to get
\begin{equation}
\label{eq:omega_derivative}
\begin{split}
	\frac{\partial \omega_k\left(\tau \right)}{\partial_{\theta_i}} &= \frac{\partial\omega_k\left(\tau-\delta\tau\right)}{\partial_{\theta_i}}+\frac{\partial\dot{\omega}_k\left(\tau-\delta\tau\right)}{\partial_{\theta_i}}\delta\tau    \\
	&=\frac{\partial \omega_k\left(0\right)}{\partial_{\theta_i}} + \sum\limits_{j=1}^{\tau/\delta\tau}\frac{\partial \dot{\omega}_k\left(j\delta\tau\right)}{\partial_{\theta_i}}\delta\tau.
\end{split}
\end{equation}	
We discuss the structure of the quantum circuits used to evaluate $\partial_{\theta_i}A$ and $\partial_{\theta_i}C$, in Appendix~\ref{app:a_c}.

In principle, the gradient of the loss function could also be approximated with a finite difference method. If the number of Hamiltonian parameters is smaller than the number of trial state parameters, this requires less evaluation circuits. However, given a trial state that has less parameters than the respective Hamiltonian, the automatic differentiation scheme presented in this section is favorable in terms of the number of evaluation circuits.
A more detailed discussion on this topic can be found in Appendix~\ref{app:complexity}.

An outline of the Gibbs state preparation and evaluation of ${\partial \omega_k\left(\tau \right)}/{\partial_{\theta_i}}$ with VarQITE is presented in Algorithm \ref{algo:VarQITE}.

\begin{algorithm}
 \caption{VarQITE for \varqbm} \label{algo:VarQITE}
\begin{algorithmic}[0]
    \State  \textbf{input}
    \State $H_{\text{eff}} = H_{\theta}^a + I^b$
	\State$\tau = {1}/{2\left(\text{k}_{\text{B}}\text{T}\right)}$
	\State $\ket{\psi\left( \omega\left(0\right)\right)} = V\left(\omega\left(0\right)\right)\ket{0}^{\otimes 2n} = \ket{\phi^{+}}^{\otimes n}$
	\State with $\ket{\phi^{+}} = \left(\ket{00}+\ket{11}\right)/{\sqrt{2}}$
    \State \textbf{procedure}
	\For{$t\in\set{\delta\tau, 2\delta\tau, \ldots, \tau}$}
		\State Evaluate $A\left(t\right)$ and $C\left(t\right)$
		\State Solve $A\dot{\omega}\left(t\right) = C$
		\For{$i\in\set{0, \ldots, p-1}$}
    			\State Evaluate $\partial_{\theta_i}C$ and $\partial_{\theta_i}A$
    			\State \label{item:linEqDer}Solve $A\left(\partial_{\theta_i}\dot{\omega}\left(t\right)\right)  = \partial_{\theta_i}C - \left(\partial_{\theta_i}A\right)\dot{\omega}\left(t\right)$
    			\State \label{item:grad_omega_tau}Compute $ \: \partial_{\theta_i}{\omega}\left(t\right) = \partial_{\theta_i}{\omega}\left(t-\delta\tau\right)+\partial_{\theta_i}\dot{\omega}\left(t\right)\delta\tau$
	   \EndFor
		\State \label{item:computeUpdateDer}Compute $\omega\left(t+\delta\tau\right) = \omega\left(t\right) +   \dot{\omega}\left(t\right)\delta\tau $
	\EndFor
	\State \textbf{return} $\omega\left(\tau\right), \: \partial\omega\left(\tau\right)/\partial\theta$
\end{algorithmic}
\end{algorithm}

Now, using a classical optimizer, such as Truncated Newton \cite{TNCDembo1983} or Adam \cite{Kingmaadam14}, allows the parameters $\theta$ to be updated according to ${\partial L}/{\partial\theta}$ from Eq.~\eqref{eq:loss_derivative}.
The \varqbm{} training is illustrated in Fig.~\ref{fig:varqbm}.

%%%%%%%%%%%%%%%%%%%%%%%%%%%%%%%%%%%%%%%%%%%%%%
\section{Results} \label{sec:results}
%%%%%%%%%%%%%%%%%%%%%%%%%%%%%%%%%%%%%%%%%%%%%%%%%%%%%%%%%%%%%%%%%%%%%%%%%%

In this section, the Gibbs state preparation with VarQITE is demonstrated using numerical simulation as well as the quantum hardware provided by IBM Quantum \cite{ibmQX}. 
Furthermore, we present numerically simulated QBM training results for a generative and a discriminative learning task.
First, aspects which are relevant for the practical implementation are discussed in Sec.~\ref{sec:impRel}. 
Next, experiments of quantum Gibbs state preparation with VarQITE are shown in Sec.~\ref{sec:VarQITEResults}.
Then, we illustrate the training of a QBM with the goal to generate a state which exhibits the sampling behavior of a Bell state, see Sec.~\ref{subsec:generative}, and to classify fraudulent credit card transactions, Sec.~\ref{subsec:discriminative}. 

%%%%%%%%%%%%%%%%%%%%%%%%%%%%
\subsection{Methods}
\label{sec:impRel}
%%%%%%%%%%%%%%%%%%%%%%%%%%%%

To begin with, we discuss the choice of a suitable parameterized trial state consisting of $V\left(\omega\right)$ and $\ket{\psi_{\text{in}}}$.
Most importantly, the initial state $\ket{\psi_{\text{in}}}$ must not be an eigenstate of $V\left(\omega\right)$ as this would imply that the circuit could only act trivially onto the state.
Furthermore, the state needs to be able to represent a sufficiently accurate approximation of the target state. If we have to represent, e.g., a non-symmetric Hamiltonian, the chosen trial state needs to be able to generate non-symmetric states.
Moreover, $V\left(\omega\right)$ should not exhibit too much symmetry as this may lead to a singular $A$ which in turn causes ill-conditioning of Eq.~\eqref{eq:sle}.
Assume, e.g., that all entries of $C$ are zero and, thus, that Eq.~\eqref{eq:sle} is homogeneous. If $A$ is singular, infinitely many solutions exist and it is difficult for the algorithm to estimate which path to choose. If $A$ is non-singular, the solution is  $\dot{\omega} = 0$ and the evolution stops although we might have only reached a local extreme point. 
Another possibility to cope with ill-conditioned systems of linear equations are least-squares methods in combination with regularization schemes. We test Tikhonov regularization \cite{Tikhonov:1620560} and Lasso regularization \cite{lassoTibshirani11} with an automatic parameter evaluation based on L-curve fitting \cite{Hansen00thel-curve}, as well as an $\epsilon$-perturbation of the diagonal, i.e., $A\rightarrow A+\epsilon I$. It turns out that all regularization methods perform similarly well. 

The results discussed in this section employ Tikhonov regularization.
Furthermore, we use trial states which are parameterized by Pauli-rotation gates. Therefore, the gradients of the QBM probabilities with respect to the trial state parameters
\begin{equation*}
   \frac{ \partial  p_v^{\text{QBM}}}{\partial\omega_k\left(\tau\right)}=\frac{\partial \text{Tr}\left[\Lambda_v\rho_{\omega}^{\text{Gibbs}}\right]} {\partial\omega_k\left(\tau\right)}
\end{equation*}
can be computed using a $\pi/2-$shift method which is, e.g., described in \cite{Zoufal2019}.
All experiments employ an additional qubit $\ket{0}_{\text{add}}$ and parameter $\omega_{\text{add}}$ to circumvent a potential phase mismatch between the target $\ket{\psi_{\tau}}$ and the trained state $\ket{ \psi\left(\omega\left(\tau\right)\right)}$ \cite{VarSITEMcArdle19, Simon18TheoryVarQSim, QNGNonUnitary19Simon} by applying
\begin{equation*}
   R_Z\left(\omega_{\text{add}}\right)\ket{0}_{\text{add}}.  
\end{equation*}
Notably, the additional parameter increases the dimension of $A$ and $C$ by one.
The effective temperature, which in principle acts as a scaling factor on the Hamiltonian parameters, is set to $\left(\text{k}_{\text{B}}\text{T}\right) = 1$ in all experiments. 

%%%%%%%%%%%%%%%%%%%%%%%%%%%%%%%%
\subsection{Gibbs State Preparation with VarQITE}
\label{sec:VarQITEResults}
%%%%%%%%%%%%%%%%%%%%%%%%%%%%%%%%
\begin{figure}[h!]
\captionsetup{singlelinecheck = false, format= hang, justification=raggedright, font=footnotesize, labelsep=space}
\begin{center}
\includegraphics[width=1\linewidth]{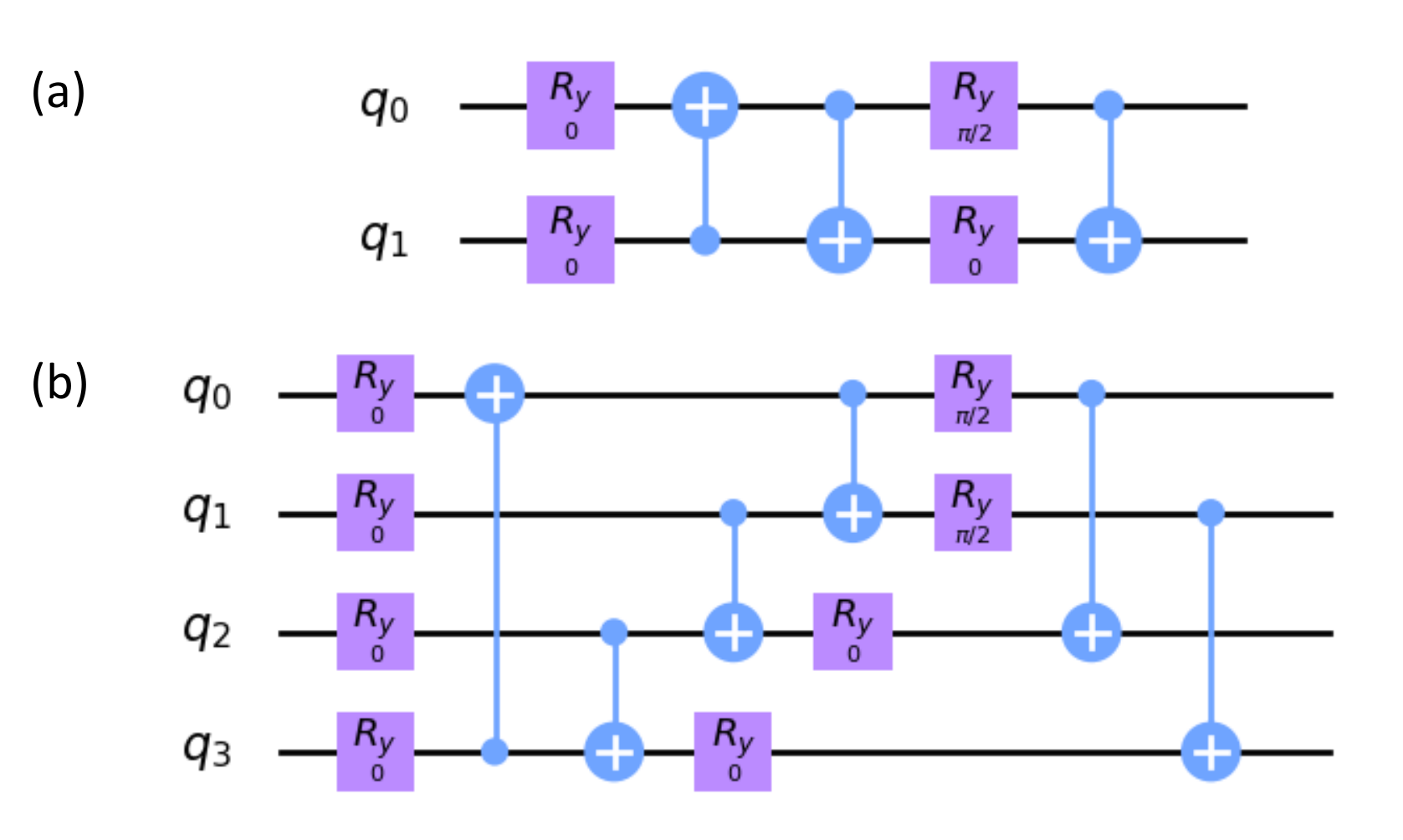}
\end{center}
\caption{The depicted circuits illustrate the initial trial state for the Gibbs state preparation of (a) $\rho^{\text{Gibbs}}_1 $ (b) $\rho^{\text{Gibbs}}_2$ using VarQITE.
}
\label{fig:ansaetze}
\end{figure}

To demonstrate that VarQITE is able to generate suitable approximations to Gibbs states, we illustrate the convergence of the state fidelity with respect to the target state for the following two simple one- and two-qubit Hamiltonians
\begin{align*}
H_1 &= 1.0 Z, \\
H_2 &= 1.0 ZZ - 0.2 ZI  - 0.2  IZ + 0.3  XI + 0.3  IX.
\end{align*}
corresponding to
\begin{align*}
\rho^{\text{Gibbs}}_1 &= \left( \begin{array}{cc}
0.12 & 0. \\
 0. & 0.88\\\end{array}\right)
, \\
\rho^{\text{Gibbs}}_2 &= \left( \begin{array}{cccc}
0.10 & -0.06 & -0.06 &  0.01 \\
-0.06 &  0.43 &  0.02 & -0.05 \\
-0.06 & 0.02 & 0.43 & -0.05 \\
 0.01 & -0.05 & -0.05 &  0.05\\\end{array}\right).
\end{align*}
The results are computed using the parameterized quantum circuit shown in Fig.~\ref{fig:ansaetze}.

\begin{figure}[h!]
\captionsetup{singlelinecheck = false, format= hang, justification=raggedright, font=footnotesize, labelsep=space}
\begin{center}
\includegraphics[width=0.8\linewidth]{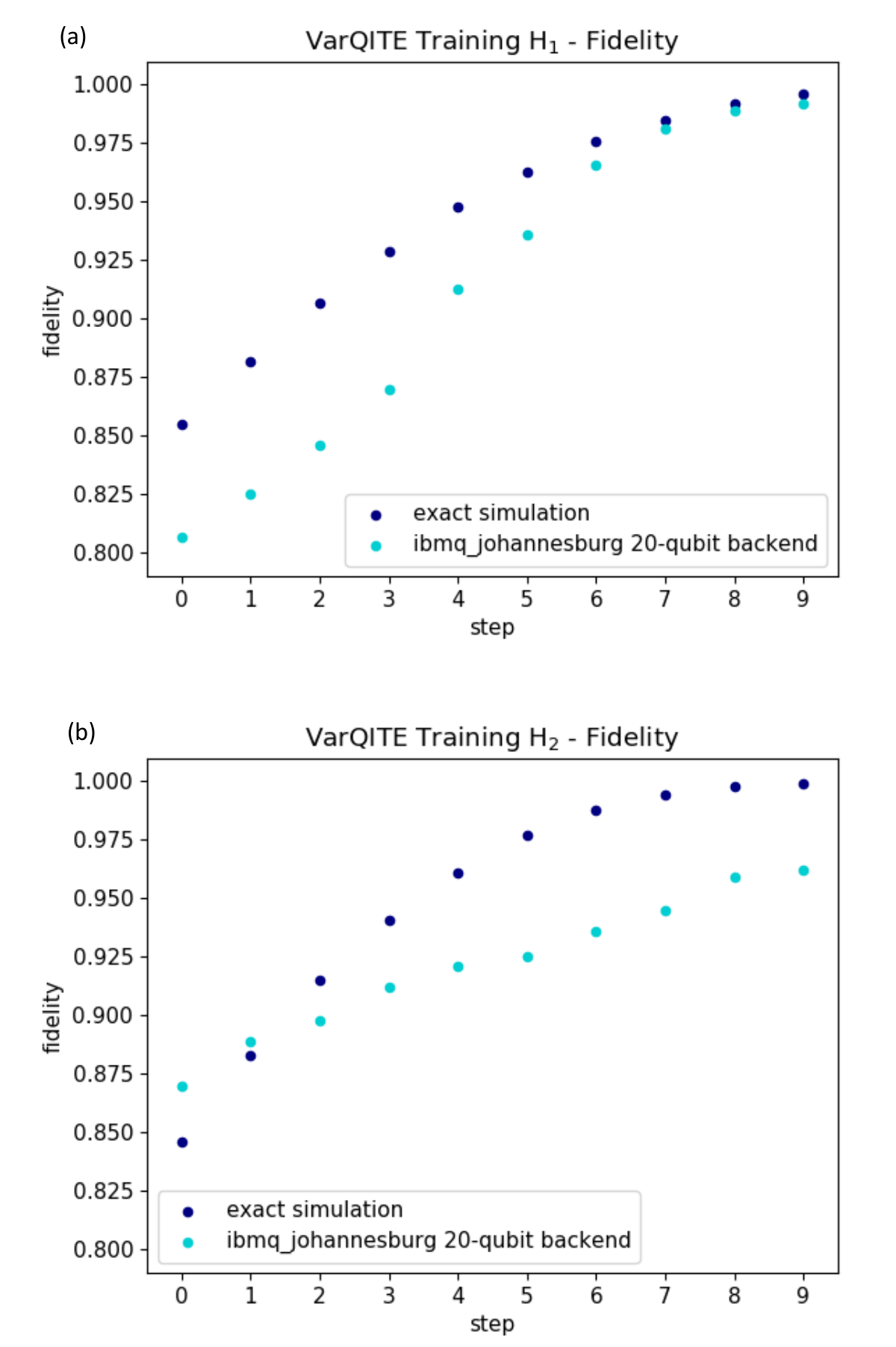}
\end{center}
\caption{Fidelity between trained and target Gibbs state with VarQITE for (a) $\rho^{\text{Gibbs}}_1$ (b) $\rho^{\text{Gibbs}}_2$ trained with an ideal simulator and real quantum hardware, i.e., the \emph{ibmq$\_$johannesburg $20$-qubit} backend. Each simulation used $10$ time steps. 
}
\label{fig:VarQITE}
\end{figure}

The algorithm is executed for $10$ time steps on different backends: an ideal simulator and the he \emph{ibmq$\_$johannesburg $20$-qubit} backend.
Notably, readout error-mitigation \cite{qiskit, dewes2012readout, Stamatopoulos2019} is used to obtain the final results run on real quantum hardware. Fig.~\ref{fig:VarQITE} depicts the results considering the fidelity between the trained and the target Gibbs state for each time step.
It should be noted that the fidelity for the quantum backend evaluations employ state tomography.
The plots illustrate that the method approximates the states, we are interested in, reasonably well and that also the real quantum hardware achieves fidelity values over $0.99$ and $0.96$, respectively.

%%%%%%%%%%%%%%%%%%%%%%%%%%%%
\subsection{Generative Learning}
\label{subsec:generative}
%%%%%%%%%%%%%%%%%%%%%%%%%%%%

\begin{figure}[h!]
\captionsetup{singlelinecheck = false, format= hang, justification=raggedright, font=footnotesize, labelsep=space}
\begin{center}
\includegraphics[width=1\linewidth]{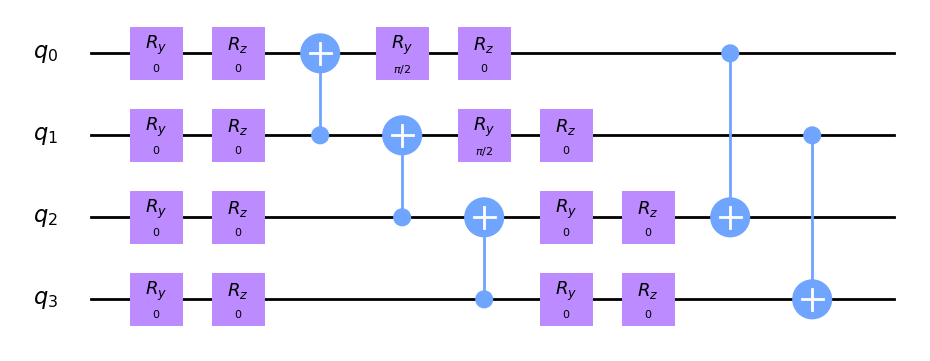}
\end{center}
\caption{We train a QBM to mimic the sampling behavior of a Bell state. 
The underlying Gibbs state preparation with VarQITE uses the illustrated parameterized quantum circuit to prepare the initial trial state. The first two qubits represent the target system and the last two qubits are ancillas needed to generate the maximally-mixed state as starting state for the evolution.}
\label{fig:ansatz_Bell}
\end{figure}

Now, the results from an illustrative example of a generative QBM model are presented.
More explicitly, the QBM is trained to mimic the sampling statistics of a Bell state $(\ket{00} + \ket{11})/\sqrt{2}$, which is a state that exhibits non-local correlations.
Numerical simulations show that the distribution can be trained with a fully visible QBM which is based on the following Hamiltonian
\begin{equation*}
	H_{\theta}= \theta_0 ZZ + \theta_1 IZ + \theta_2 ZI.
\end{equation*}
We draw the initial values of the Hamiltonian parameters $\theta$ from a uniform distribution on $\left[-1, 1\right]$.
The optimization runs on an ideal simulation of a quantum computer using AMSGrad \cite{amsgrad} with initial learning rate $0.1$, maximum number of iterations $200$, first momentum $0.7$, and second momentum $0.99$ as optimization routine.
The Gibbs state preparation uses the initial trial state shown in Fig.~\ref{fig:ansatz_Bell} and $10$ steps per state preparation.

\begin{figure}[h!]
\captionsetup{singlelinecheck = false, format= hang, justification=raggedright, font=footnotesize, labelsep=space}
\begin{center}
\includegraphics[width=0.8\linewidth]{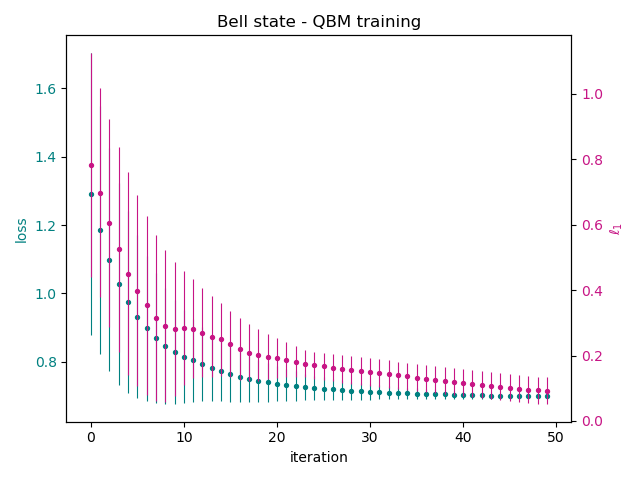}
\end{center}
\caption{The figure illustrates the training progress of a fully-visible QBM model which aims to represent the measurement distribution of a Bell state.
The green function corresponds to the loss and the pink function represents the distance between the trained and target distribution with respect to the $\ell_1$ norm at each step of the iteration. 
Both measures are computed for $10$ different random seeds. 
The points represent the mean and the error bars the standard deviation of the results.
}
\label{fig:qrbm}
\end{figure}

\begin{figure}[h!]
\captionsetup{singlelinecheck = false, format= hang, justification=raggedright, font=footnotesize, labelsep=space}
\begin{center}
\includegraphics[width=0.8\linewidth]{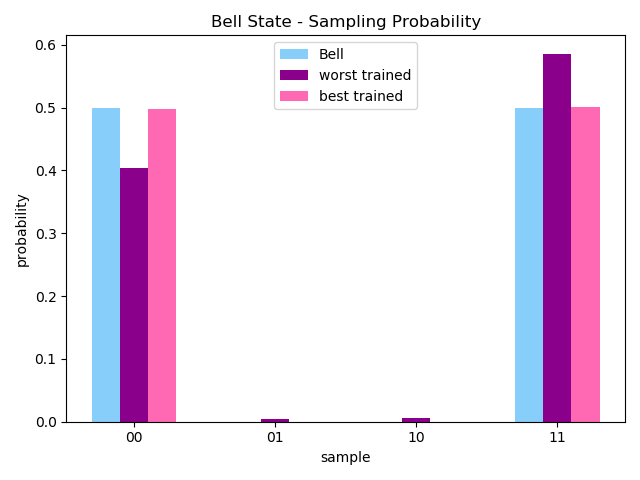}
\end{center}
\caption{The figure illustrates the sampling probability of the Bell state (blue), as well as the best (pink) and worst (purple) probability distribution achieved from  $10$ different random seeds. 
}
\label{fig:bell_prob}
\end{figure}

The training is run $10$ times using different randomly drawn initial parameters.
The averaged values of the loss function as well as the distance between the target distribution $p^{\text{data}}=\left[0.5, 0., 0., 0.5\right]$ and the trained distribution $p^{\text{QBM}}$ with respect to the $\ell_1$ norm are illustrated over $50$ optimization iterations in Fig.~\ref{fig:qrbm}. 
The plot shows that loss and distance converge toward the same values for all sets of initial parameters. Likewise, the trained parameters $\theta$ converge to similar values.
Furthermore, Fig.~\ref{fig:bell_prob} illustrates the target probability distribution and for the best and worst of the trained distributions. The plot reveals that the model is able to train the respective distribution very well.

%%%%%%%%%%%%%%%%%%%%%%%%%%%%
\subsection{Discriminative Learning}
\label{subsec:discriminative}
%%%%%%%%%%%%%%%%%%%%%%%%%%%%

QBMs are not only applicable for generative but also for discriminative learning. We discuss the application to a classification task, the identification of fraudulent credit card transactions.

To enable discriminative learning with QBMs, we use the input data points $x$ as bias for the Hamiltonian weights. More explicitly, the parameters of the Hamiltonian
\begin{equation}
    H_{\theta}\left(x\right)=\sum\limits_{i}f_i\left(\theta, x\right)h_i
\end{equation}
are given by a function $f_i\left(\theta, x\right)$ which maps $\theta$ and $x$ to a scalar in $\mathbb{R}$.
Now, the respective loss function reads
\begin{equation*}
\label{eq:loss_sup}
\begin{split}
	L = -\sum\limits_{x}p_x^{\text{data}}\sum\limits_{v}p_{v|x}^{\text{data}}\log{p_{v|x}^{\text{QBM}}}
\end{split}
\end{equation*}
with
\begin{equation*}
   p_{v|x}^{\text{QBM}} = \text{Tr}\left[\Lambda_v\rho\left(x\right)_{\omega}^{\text{Gibbs}}\right],
\end{equation*}
where $\rho\left(x\right)_{\omega}^{\text{Gibbs}}$ denotes the approximate Gibbs state corresponding to $H_{\theta}\left(x\right)$.
The model encodes the class labels in the measured output configuration of the visible qubits $v$ of $\rho\left(x\right)_{\omega}^{\text{Gibbs}}$.
Now, the aim of the training is to find Hamiltonian parameters $\theta$ such that, given a data sample $x$, the probability of sampling the correct output label from $\rho\left(x\right)_{\omega}^{\text{Gibbs}}$ is maximized.

The training is based on $500$ artificially created credit card transactions \cite{Altman2019} with about $15\%$ fraudulent instances. To avoid redundant state preparation, the training is run for all unique item instances in the data set and the results are averaged according to the item's occurrence counts.
The dataset includes the following features: location (ZIP code), time, amount, and Merchant Category Code (MCC) of the transactions.
To facilitate the training, the features of the given data set are discretized and normalized as follows. Using k-means clustering, each of the first three features are independently discretized to $3$ reasonable bins. Furthermore, we consider MCCs $<10 000$ and group them into $10$ different categories.
The discretization is discussed in more detail in Table \ref{tbl:discr_data_preproc}.
Furthermore, for each feature, we map the values $x$ to $x' = \frac{x-\mu}{\sigma}$ with $\mu$ denoting the mean and $\sigma$ denoting the standard deviation. 
\begin{table}[h!]
\captionsetup{singlelinecheck = false, format= hang, justification=raggedright, font=footnotesize, labelsep=space}
%\begin{tabular}{|l|l|l|}
\begin{tabular}{c|c|c}
Feature & Condition & Value \\
 \hline
\multirow{2}{*}{Time} & $0$AM $- 11$AM  & 0 \\
& $11$AM $- 6$PM  & 1 \\
& $6$PM - $0$AM & 2 \\
\hline
\multirow{2}{*}{Amount} & amount < $\$50 $ & 0 \\
& amount in $\$50-150 $ & 1 \\ 
& amount > $\$150$ & 2 \\
\hline
\multirow{3}{*}{ZIP} & east & 0 \\ 
& central & 1 \\ 
& west & 2 \\
\end{tabular}
\caption{The table discusses the clustering of a transaction fraud data set which is used to train a discriminative QBM model. MCC refers to the merchant category code and ZIP to zone improvement plan.
Notably, the given values are approximate.}
\label{tbl:discr_data_preproc}
\end{table}

The complexity of this model demands a Hamiltonian that has sufficient representation capabilities. Our choice is the following
\begin{equation}
\label{eq:H_disc}
\begin{split}
    H_{\theta}\left(x\right) =&\:f_0\left(\theta, x\right)  ZZ + f_1\left(\theta, x\right)ZI  +\\
    &\:f_2\left(\theta, x\right) IZ + f_3\left(\theta, x\right)XI + f_4\left(\theta, x\right) IX,
\end{split}
\end{equation}
where $f_i\left(\theta, x\right) = \vec{\theta}_i\cdot\vec{x}$ corresponds to the dot product of the vector corresponding to the data item $\vec{x}$ and a parameter vector $\vec{\theta}_i$ of equal length. Additionally, the first and second qubit correspond to a hidden and visible qubit, respectively.

Since the numerical simulation of variational Gibbs state preparation for various $H_{\theta}\left( x\right)$, with $x$ corresponding to all unique data items, is computationally expensive, we decided to train the parameters $\theta$ using an exact representation of the quantum Gibbs states.
The resulting $\theta$ are then used for Gibbs state preparation with VarQITE.
Even though the parameters are not trained with variational Gibbs state preparation, the results discussed in this section demonstrate that we can find suitable parameters $\theta$ such that \varqbm{} corresponds to a well-performing discriminative model.

The exact training uses a Truncated Netwon optimization routine \cite{TNCDembo1983} with a maximum iteration number of $100$ and the step size for the numerical approximation of the Jacobian being set to $10^{-6}$. 
The initial values for the Hamiltonian parameters are drawn from a uniform distribution on $\left[-1, 1\right]$.
 \begin{figure}[h!]
\captionsetup{singlelinecheck = false, format= hang, justification=raggedright, font=footnotesize, labelsep=space}
\begin{center}
\includegraphics[width=1\linewidth]{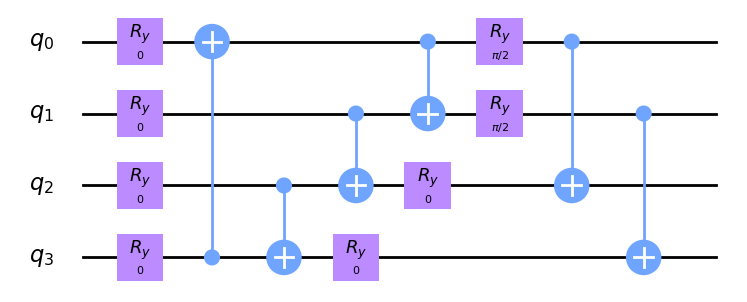}
\end{center}
\caption{Given a transaction instance, the measurement output of the QBM labels it as being either fraudulent or valid. The underlying Gibbs state preparation with VarQITE uses the illustrated parameterized quantum circuit as initial trial state. The first qubit is the visible node that determines the QBM output, the second qubit represents the hidden unit, and the last two qubits are ancillas needed to generate the maximally-mixed state as starting state for the evolution.}
\label{fig:ansatz_Disc}
\end{figure}

Given a test data set consisting of $250$ instances, with about $10\%$ fraudulent transactions, the Gibbs states, corresponding to the unique items of the test data, are approximated using VarQITE with the trained parameters $\theta$ and the trial state shown in Fig.~\ref{fig:ansatz_Disc}. 
To predict the labels of the data instances, we sample from the states $\rho_{\omega}^{\text{Gibbs}}$ and choose the label with the highest sampling probability.
These results are, then, used to evaluate the accuracy, precision, recall and F$_1$ score.
It should be noted that we choose a relatively simple quantum circuit to keep the simulation cost small. 
However, it can be expected that a more complex parameterized quantum circuit would lead to further improvement in the training results.

The resulting values are compared to a set of standard classifiers defined in a \emph{scikit-learn} \cite{scikit-learn2011} classifier comparison tutorial \cite{scikitClassifierComp}, see Tbl.~\ref{tbl:measuresQBM}. The respective classifiers are used with the hyper parameters defined in this tutorial.
Notably, the Linear SVM does not classify any test data item as fraudulent and, thus, the classifier sets precision and recall score to $0$.
The comparison reveals that the QBM performs similarly well to the classical classifiers considering accuracy, is competitive regarding precision, and even outperforms them in terms of recall. The best F$_1$ score is achieved with \varqbm.

\begin{table}[h!]
\captionsetup{singlelinecheck = false, format= hang, justification=raggedright, font=footnotesize, labelsep=space}
{\renewcommand{\arraystretch}{1.2}
\begin{tabular}{c | c | c | c | c}
Model & Accuracy & Recall & Precision & F$_1$\\
\hline
Nearest Neighbours & $0.94$ &  $0.54$ & $0.72$ & $0.31$ \\
Linear SVM & $0.90 $& $0$ & $0$ & $0$ \\
RBF SVM & $0.94 $& $0.42$ & $0.83$ & $0.28$  \\
Gaussian Process & $0.94$ & $0.46$ & $0.85$ & $0.30$\\
Gaussian Naive Bayes  & $0.91$ &$ 0.42 $& $0.56$ & $0.24$\\
Decision Tree & $0.94$ & $0.42$ & $0.83$ & $0.28$\\
Random Forrest & $0.93 $& $0.29$ & $1.00$ & $0.22$\\
Multi-layer Perceptron  & $0.94$ &  $0.38$ &$0.9$ & $0.27$\\
AdaBoost & $0.94$ & $0.54$ & $0.81$ & $0.32$\\
QDA & $0.92$ & $0.46$ & $0.61$ & $0.26$\\
\textbf{\varqbm} & $\mathbf{0.95}$ & 
$\mathbf{0.63}$ & $\mathbf{0.83}$ & $\mathbf{0.36}$\\
\end{tabular}
}
\caption{This table presents performance measures for scikit-learn standard classifiers, as well as the trained QBM. The Nearest Neighbours classifier uses a $3$ nearest neighbours vote. The Linear and RBF Support Vector Machine (SVM) are based on a linear and radial kernel, respectively. The Linear SVM uses a regularization term of $0.25$ and for the RBF SVM the kernel coefficient is set to $2$. The maximum depth of the Decision Tree as well as the Random Forrest is set to 5. Furthermore, the Random Forrest classifier uses $10$ trees and uses $1$ feature to search for the best spit. The Multi-layer Perceptron uses $\ell_2$ regularization with coefficient $1$ and a maximum iteration number of $1000$. QDA refers to Quadratic Discriminant Analysis. It should be noted that the remaining classifier properties are default settings.}
\label{tbl:measuresQBM}
\end{table}

%%%%%%%%%%%%%%%%%%%%%%%%%%%%%%%%%%%%%%%%%%%%%%%%%%%%%%%%%%%%%%%%%%%%%%%%%%}
\section{Conclusion and Outlook}
\label{sec:discussion}
%%%%%%%%%%%%%%%%%%%%%%%%%%%%%%%%%%%%%%%%%%%%%%%%%%%%%%%%%%%%%%%%%%%%%%%%%%

This work presents the application of McLachlan's variational principle to facilitate \varqbm, a variational QBM algorithm, that is compatible with generic Hamiltonians and can be trained using analytic gradients of the actual loss function even if some of the qubits are hidden. Suppose a sufficiently powerful variational trial state, the presented scheme is not only compatible with local but also long-range correlations and for arbitrary system temperatures.

We outline the practical steps for utilizing VarQITE for Gibbs state preparation and verify that it can train states which are reasonably close to the target using simulation as well as real quantum hardware.
Moreover, applications to generative learning and classification are discussed and illustrated with further numerical results. 
The presented model offers a versatile framework which facilitates the representation of complex structures with quantum circuits.

An interesting question for future research is the investigation of performance measures that improve our understanding of the model's representation capabilities. 
Furthermore, QBMs are not limited to the presented applications. They could also be utilized to train models for data from experiments with quantum systems. This is a problem that has recently gained interest, see e.g., \cite{LearningModels20}.  Additionally, they might be employed for combinatorial optimization. Classical BMs have been investigated in this context \cite{Spieksma1995} and developing and analyzing quantum algorithms for combinatorial optimization is an active area of research \cite{FarhiQAOA14, Barkoutsos20VQOCVaR}. 

All in all, there are many possible applications which still have to be explored.

%%%%%%%%%%%%%%%%%%%%%%%%%%%%%%%%%%%%%%%%%%%%%%%%%%%%%%%%%%%%%%%%%%%%%%%%%%
\section{Acknowledgments}
%%%%%%%%%%%%%%%%%%%%%%%%%%%%%%%%%%%%%%%%%%%%%%%%%%%%%%%%%%%%%%%%%%%%%%%%%%

We would like to thank Erik Altman for making the synthetic credit card transaction dataset available to us. 
Moreover, we are grateful to Pauline Ollitrault, Guglielmo Mazzola and Mario Motta for sharing their knowledge and engaging in helpful discussion.
Furthermore, we thank Julien Gacon for his help with the implementation of the algorithm and all of the IBM Quantum team for its constant support.

Also, we acknowledge the support of the National Centre of Competence in Research \textit{Quantum Science and Technology} (QSIT).

IBM, the IBM logo, and ibm.com are trademarks of International Business Machines Corp., registered in many jurisdictions worldwide. Other product and service names might be trademarks of IBM or other companies. The current list of IBM trademarks is available at \url{https://www.ibm.com/legal/copytrade}.

%%%%%%%%%%%%%%%%%%%%%%%%%%%%%%%%%%%%%%%%%%%%%%%%%%%%%%%%%%%%%%%%%%%%%%%%%%%
%\section{Author Contributions}
%%%%%%%%%%%%%%%%%%%%%%%%%%%%%%%%%%%%%%%%%%%%%%%%%%%%%%%%%%%%%%%%%%%%%%%%%%%
%All authors researched, collated, and wrote this paper.
%
%%%%%%%%%%%%%%%%%%%%%%%%%%%%%%%%%%%%%%%%%%%%%%%%%%%%%%%%%%%%%%%%%%%%%%%%%%%
%\section{Competing Interests}
%%%%%%%%%%%%%%%%%%%%%%%%%%%%%%%%%%%%%%%%%%%%%%%%%%%%%%%%%%%%%%%%%%%%%%%%%%%
%The authors declare that there are no competing interests.
%
%%%%%%%%%%%%%%%%%%%%%%%%%%%%%%%%%%%%%%%%%%%%%%%%%%%%%%%%%%%%%%%%%%%%%%%%%%%
%\section{Data Availability}
%%%%%%%%%%%%%%%%%%%%%%%%%%%%%%%%%%%%%%%%%%%%%%%%%%%%%%%%%%%%%%%%%%%%%%%%%%%
%The data that support the findings of this study are available from the corresponding author upon reasonable request.

\appendix

%%%%%%%%%%%%%%%%%%%%%%%%%%%%%%%%%%%%%%%%%%%%%%%%%%%%%%%%%%%%%%%%%%%%%%%%%%
\section{Evaluation of A, C and their gradients}
\label{app:a_c}
%%%%%%%%%%%%%%%%%%%%%%%%%%%%%%%%%%%%%%%%%%%%%%%%%%%%%%%%%%%%%%%%%%%%%

\begin{figure}[h!]
\captionsetup{singlelinecheck = false, format= hang, justification=raggedright, font=footnotesize, labelsep=space}
\begin{center}
\includegraphics[width=1\linewidth]{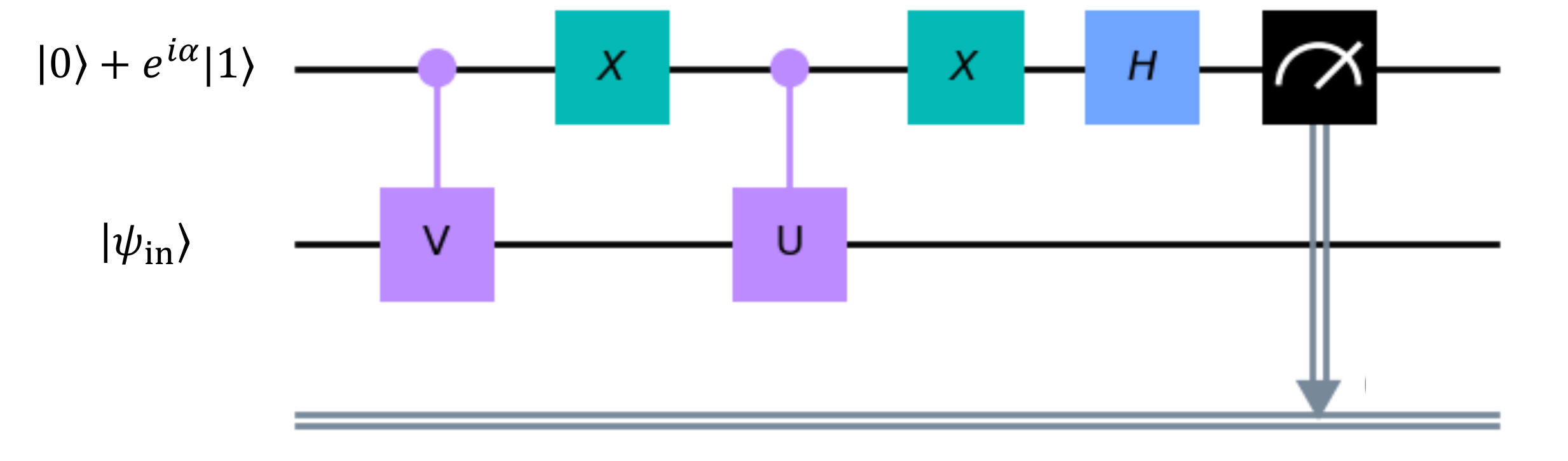}
\end{center}
\caption{Quantum circuit to evaluate $\text{Re}\left(e^{i\alpha}\text{Tr}\left[U^{\dagger}V\rho_{\text{in}}\right]\right)$ with $\rho_{\text{in}} = \ket{\psi}\bra{\psi_{\text{in}}}$.}
\label{fig:expValueCircuit}
\end{figure}

The elements of the matrix $A$ and the vector $C$, see Eq.~\eqref{eq:AC} are of the following form
\begin{equation}
\label{eq:expValueEval}
	\text{Re}\left(e^{i\alpha}\text{Tr}\left[U^{\dagger}V\rho_{\text{in}}\right]\right)
\end{equation}
with $\text{Re}\left(\cdot\right)$ denoting the real part and $\rho_{\text{in}} = \ket{\psi_{\text{in}}}\bra{\psi_{\text{in}}}$.
As discussed in \cite{VarSITEMcArdle19, Simon18TheoryVarQSim, LaflammeSimulatingPhysPhenom02}, such terms can be computed by sampling the expectation value of an observable $Z$ with respect to the quantum circuit shown in Fig.~\ref{fig:expValueCircuit}.

Notably, the phase $e^{i\alpha}$ in the first qubit is needed to include phases which may occur from gate derivatives. In our case, $\alpha$ needs to be set to $0$ respectively $\pi/2$ when computing the terms of $A$ or $C$. More precisely, the first qubit is initialized by an $H$ gate for $A$ and $H$ followed by an $S$ gate for $C$.
These phases come from the fact that the trial states, used in this work, are constructed via Pauli rotations, i.e., $U\left(\omega\right) = R_{\sigma_l}\left(\omega\right)$ with $\sigma_l \in \set{X, Y, Z}$, which leads to
\begin{equation}
 \frac{\partial U\left(\omega\right)}{\partial\omega}  = -\frac{i}{2}{\sigma_l}R_{\sigma_l}\left(\omega\right).
\end{equation}
Furthermore, this method can be applied for the evaluation of $\partial A / \partial \theta$ and $\partial C / \partial \theta$, i.e., the respective terms can be written in the form of Eq.~\eqref{eq:expValueEval}.
More precisely,
\begin{equation*}
\label{eq:dH_A}
\begin{split}
	&\partial_{\theta_i}A_{p,q}\left(\tau\right) =\\
&\sum\limits_s \frac{\partial\omega_s\left(\tau\right)}{\partial_{\theta_i}}\text{Re}\left(\text{Tr}\left[\left(\frac{\partial^2 V^{\dagger}\left({\omega\left(\tau\right)}\right)}{\partial\omega_p\left(\tau\right) \partial\omega_s\left(\tau\right)}\frac{\partial V\left({\omega\left(\tau\right)}\right)}{\partial\omega\left(\tau\right)_q}\right. \right.\right.\\
&\left.\left.\left. + \frac{\partial V^{\dagger}\left({\omega\left(\tau\right)}\right)}{\partial\omega\left(\tau\right)_p}\frac{\partial^2 V\left({\omega\left(\tau\right)}\right)}{\partial\omega_q\left(\tau\right)\partial\omega_s\left(\tau\right)}\right)\rho_{\text{in}}\right]\right)
\end{split}
\end{equation*}
and
\begin{equation*}
\label{eq:dH_C}
\begin{split}
	&\partial_{\theta_j}C_p = \\
	&-\text{Re}\left(\text{Tr}\left[\frac{\partial V^{\dagger}\left({\omega\left(\tau\right)}\right)}{\partial\omega\left(\tau\right)_p}h_jV\left({\omega\left(\tau\right)}\right)\rho_{\text{in}}\right]\right)\\
&\left. - \sum\limits_{i,s}\theta_i\frac{\partial\omega_s\left(\tau\right)}{\partial_{\theta_j}}\text{Re}\left( \text{Tr}\left[\left(\frac{\partial V^{\dagger}\left({\omega\left(\tau\right)}\right)}{\partial\omega_p\left(\tau\right)}h_i\frac{\partial V\left({\omega\left(\tau\right)}\right)}{\partial\omega_s\left(\tau\right)} \right.\right. \right.\right.\\
 &+ \left.\left. \left.\frac{\partial^2 V^{\dagger}\left({\omega\left(\tau\right)}\right)}{\partial\omega_p\left(\tau\right)\partial\omega_s\left(\tau\right)} h_iV\left({\omega\left(\tau\right)}\right)\right)\rho_{\text{in}} \right] \right).
	\end{split}
\end{equation*}
Hereby, $\alpha$ must be set to $\pi/2$ respectively $0$ for the terms in $\partial A/ \partial\theta$ respectively $\partial C/ \partial\theta$. This is achieved with the same gates as mentioned before.

%%%%%%%%%%%%%%%%%%%%%%%%%%%%%%%%%%%%%%%%%%%%%%%%%%%%%%%%%%%%%%%%%%%%%
\section{Complexity Analysis}
\label{app:complexity}
%%%%%%%%%%%%%%%%%%%%%%%%%%%%%%%%%%%%%%%%%%%%%%%%%%%%%%%%%%%%%%%%%%%%%

To compute the gradient $ \partial L / \partial \theta$ of the loss function, given in Eq.~\eqref{eq:loss_qbm}, we could use either a numerical finite differences method \cite{Kardestuncer1975}, or the analytic, automatic differentiation approach that is presented in this paper. In the following, we discuss the number of circuits that have to be evaluated for those gradient implementations for a trial state with $q$ parameters, an $n$-qubit Hamiltonian with $p$ parameters, and VarQITE for Gibbs state preparation using $t$ steps.

The number of circuits that need to be evaluated for Gibbs state preparation with VarQITE are $\Theta\left(tq^2 \right)$ and $\Theta\left(tqp \right)$ for $A$ and $C$, respectively.
Therefore, the overall number of circuits is $\Theta\left(tq(q+p)\right)$.
Now, computing the gradient with forward finite differences reads
\begin{equation*}
    \frac{\partial L}{\partial \theta} \approx  \frac{L\left( \theta + \epsilon \right) - L\left( \theta  \right)}{\epsilon},
\end{equation*}
for $0 < \epsilon \ll 1$. 
For this purpose, VarQITE must be run once with $\theta$ and $p$ times with an $\epsilon$-shift which leads to a total number of $\Theta\left(tpq(q+p)\right)$ circuits.

The automatic differentiation gradient, given in Eq.~\eqref{eq:loss_derivative}, corresponds to
\begin{equation*}
\begin{split}
	\frac{\partial L}{\partial\theta} = - \sum\limits_{v}\sum\limits_{k=0}^{q-1}\frac{p_v^{\text{data}}}{\left\langle \Lambda_v\right\rangle} \frac{\partial \left\langle \Lambda_v \right\rangle}{\partial \omega_k} \frac{\partial \omega_k} {\partial \theta}
		\end{split}
\end{equation*}
with $\langle \ldots \rangle = \text{Tr}\left[\rho_{\omega}^{Gibbs}\ldots\right]$.
VarQITE needs to be run once to prepare $\rho_{\omega}^{\text{Gibbs}}$. Furthermore, the evaluation of $\partial\omega_k / \partial\theta$ requires that $\partial A / \partial\theta$ and $\partial C / \partial\theta$ are computed for every step of the Gibbs state preparation. This leads to $\Theta\left(tq^2(q+p)\right)$ circuits.
The resulting overall complexity of the number of circuits is $\Theta\left(tq^2(q+p)\right)$.

The results are summarized in Tbl.~\ref{tbl:complexity}.
Automatic differentiation is more efficient than finite differences if $q < p$. For $q > p$, on the other hand, focusing mainly on computational complexity, one should rather use finite differences. 
Considering, e.g., a $k$-local Ising model that corresponds to a Hamiltonian with $\mathcal{O}\left(n^k\right)$ parameters. Suppose that we can find a reasonable variational $n$-qubit trial state with $\mathcal{O}\left(n\right)$ layers of parameterized and entangling gates, which results in $q = \mathcal{O}\left(n^2\right)$ parameters, then, automatic differentiation would outperform finite differences for $k>2$.

\begin{table}[h!]
\captionsetup{singlelinecheck = false, format= hang, justification=raggedright, font=footnotesize, labelsep=space}
{\renewcommand{\arraystretch}{1.2}
\begin{tabular}{ c | c }
Method & Number Circuits \\ 
\hline
Finite Diff &  $\Theta\left(tqp(q+p)\right)$\\
Automatic Diff   & $\Theta\left(tq^2(q+p)\right)$\\
\end{tabular}
}
\caption{Comparing the number of circuits needed to train a QBM with VarQITE using either finite differences or automatic differentiation. The number of Hamiltonian parameters is $p$, the number of trial state parameters is $q$ and the number of time steps during the Gibbs state preparation is $t$.}
\label{tbl:complexity}
\end{table}

\bibliographystyle{IEEEtranN}
\bibliography{references}

\end{document}